# Soft x-ray ptychography with SOPHIE: Guide and instrumentation

Tim A. Butcher; Simone Finizio; Lars Heller; Nicholas W. Phillips; Blagoj Sarafimov; Carlos A. F. Vaz; Armin Kleibert; Benjamin Watts; Mirko Holler; Jörg Raabe

# Soft x-ray ptychography with SOPHIE: Guide and instrumentation

Tim A. Butcher,[1,2,a)] Simone Finizio,[1] Lars Heller,[1] Nicholas W. Phillips,[1,3] Blagoj Sarafimov,[1] Carlos A. F. Vaz,[1] Armin Kleibert,[1] Benjamin Watts,[1] Mirko Holler,[1] and Jörg Raabe[1,a)]

## AFFILIATIONS

[1] Paul Scherrer Institut, 5232 Villigen PSI, Switzerland
[2] Max Born Institute for Nonlinear Optics and Short Pulse Spectroscopy, 12489 Berlin, Germany
[3] Mineral Resources CSIRO, 3168 Clayton, Australia

[a)] **Authors to whom correspondence should be addressed:** tim.butcher@psi.ch and joerg.raabe@psi.ch

## ABSTRACT

Soft x-ray ptychography is becoming a key synchrotron microscopy technique in the fields of condensed matter physics, chemistry, and environmental and life sciences. Its attractiveness across broad disciplinary fields is owed to the favorable combination of high spatial resolution and strong contrast mechanisms. The SOft X-ray Ptychography Highly Integrated Endstation (SOPHIE) at the Swiss Light Source (SLS) was developed to accommodate soft x-ray ptychography experiments requiring high spatial resolution, in addition to high chemical and ferroic sensitivities. An introduction to soft x-ray ptychography with SOPHIE aimed at prospective users is provided. Furthermore, an overview of the instrumentation of SOPHIE is given along with an example of the imaging capabilities, which demonstrate the achievement of a sub-10 nm spatial resolution at a photon energy of 706 eV.

## I. INTRODUCTION

The first demonstrations and development of coherent diffractive imaging (CDI) techniques for microscopy with synchrotron radiation began in the late 1990s and continued in the new millennium.[1–3] This approach consists of recording a diffraction pattern obtained by scattering of a coherent x-ray beam from a sample and subsequent numerical conversion to an image by solution of the phase problem.[4,5] A major advantage of CDI over conventional imaging is that the spatial resolution is not limited by the ability to focus the radiation, which is challenging in the x-ray regime. Of the CDI methods, x-ray ptychography in particular is establishing itself as a microscopy method at synchrotrons.[6,7]

Ptychographic imaging relies on the collection of coherent diffraction patterns by a two-dimensional pixelated detector with a high dynamic range for a set of overlapping sample positions. Although first proposed by Walter Hoppe for electron microscopy in 1968,[8–10] technological limitations in detection, computation, and scanning precision meant that ptychography was impractical at that time and was shelved until its revival in the early 2000s.[11–13] X-ray ptychography became feasible with the advent of high coherent flux at third generation synchrotron sources with fast computational resources and is now widespread at facilities around the world.

Compared to hard x-ray imaging, experiments in the soft x-ray energy window of 200–2000 eV (6.2–0.62 nm wavelength) are particularly challenging concerning photon detection and the requirement of high vacuum to avoid attenuation of the x-ray beam. Severe absorption of soft x-rays in matter also limits the maximum sample thicknesses to a couple of hundred nanometers. Microscopy with soft x-rays is nonetheless valuable due to contrast mechanisms that enable microspectroscopy with high chemical and ferroic sensitivities. This is made possible by tuning the photon energy to the x-ray absorption edges of relevant elements. Thus, chemical characterization is possible at the nanoscale in the form of spatially resolved x-ray absorption spectroscopy (XAS).[14] Furthermore, magnetic information can be obtained from the detected intensities in the form of x-ray magnetic circular dichroism (XMCD) maps with circularly polarized x-rays.[15] Ferroelectric and antiferromag-netic order can also be accessed with x-ray linear dichroism (XLD) or x-ray magnetic linear dichroism (XMLD) by variation of the linear polarization angle. For most elements, absorption edges in the hard x-ray regime only probe excited states with strongly reduced sensitivities to the chemical environment and ferroic ordering.

The absorption edge at the transition from 1s → 2p electronic states, termed the K-edge, includes light elements such as C (290 eV), N (400 eV), O (530 eV), and F (685 eV) in the soft x-ray range.[16] Details about the chemical environment can be extracted due to high sensitivity to spatial distortions of the 2p states and hybridizations with the electron orbitals of other elements. Samples with 3d transition metals are well characterized by the 2p → 3d process of the L-edge ($L_3$-edges: Cr: 575 eV, Mn: 640 eV, Fe: 708 eV, Co: 778 eV, Ni: 855 eV, and Cu: 931 eV).[16] The lanthanides that are commonly referred to as rare earth metals and have near identical chemical properties due to the localized 4f electrons are straightforward to distinguish at their M-edges 3d → 4f. In the case of the actinides, the N-edge with the 4d → 5f transition can provide important information.

In recent years, synchrotron-based x-ray ptychography has been highly successful by combining elements from CDI and scanning transmission x-ray microscopy (STXM). In STXM measurements, x-rays are focused by a Fresnel Zone Plate (FZP) to a small spot in the focal plane through which the sample is moved.[17,18] Successive image pixels are measured by raster-scanning the sample through the beam focus while detecting the transmitted photons with a point detector such as an avalanche photodiode (APD) or a phosphor-coupled photomultiplier. However, the spatial resolution of STXM is limited by the focal spot size of the FZP, which in turn is dictated by the lithographic procedure for the fabrication of the thin, high aspect ratio, outermost zone walls. The spatial resolution of STXM is around 25 nm for routine operation. An improvement of the spatial resolution to sub-15 nm with specialized FZPs is possible, but involves limitations due to challenging FZP fabrication, short focal distances, and low efficiency of the FZPs.[19] It is important to note that STXM also relies on the coherence of the x-ray beam incident on the FZP, as incoherent illumination of the diffractive optic leads to a larger spot size and loss of spatial resolution. Fully coherent illumination of the FZP provides a spot size of ~1.2 times the outermost zone width.[18]

X-ray ptychography typically uses a similar optical setup as STXM, except that the point detector is replaced by a pixelated area detector. In terms of observed contrast, STXM measures a point in the transmission XAS spectrum (imaginary part of the refractive index) as a function of position, while CDI methods such as x-ray ptychography measure the intensity of coherent diffraction patterns from which complex-valued images are reconstructed. Both real and imaginary parts of the refractive index can be extracted from the reconstructed data.

A significant factor limiting the spatial resolution of coherent diffractive imaging techniques such as x-ray ptychography is the ability to detect the high spatial frequencies of the diffracted x-rays. This requires that the detector covers a wide angular area, which is typically achieved by placing the flat detector close to the sample, and that the detector has low noise so that the sparse scattering events at the widest angles are significantly above the noise floor. In contrast to other CDI techniques, x-ray ptychography makes use of overlapping illuminated areas between adjacent scan points when solving the phase problem. Ptychographic imaging has fewer requirements for the sample as in, for example, Fourier transform holography that is comparatively robust to drifts of the sample but relies on the preparation of masks with reference holes that enable phase retrieval.[20–26] In addition to the facilitated sample preparation, x-ray ptychography has few impediments to the imaging of two-dimensionally extended samples.[27,28]

Notable imaging activity with soft x-ray ptychography has been carried out at the COSMIC beamline of the Advanced Light Source (ALS, Lawrence Berkeley National Laboratory, USA),[28–35] at I08 of Diamond light source (Didcot, United Kingdom),[36,37] Hermes at Soleil (Saint-Aubin, France),[38–41] the 10ID-1 Soft X-ray Spectromicroscopy beamline at the Canadian Light Source (Saskatoon, Canada),[42–45] the SIM beamline at the Swiss Light Source (SLS, Villigen, Switzerland),[46,47] the SoftiMAX beamline at MAX IV (Lund, Sweden),[48,49] the CARROT endstation at the BL07LSU beamline of SPring-8 (Sayō, Japan),[50,51] the SSRF–BL08U1A beamline at Shanghai Synchrotron Radiation Facility (China),[52,53] BESSY II (Berlin, Germany),[54–58] TwinMic at Elettra (Basovizza, Italy),[59,60] and early work at the 2-ID-B beamline of the Advanced Photon Source (Argonne National Laboratory, USA).[61–63]

The present report provides a description of the SOft X-ray Ptychography Highly Integrated Endstation (SOPHIE) and a demonstration of its imaging capabilities. The SOPHIE microscope was designed to be adaptable and mobile, enabling different sample environments, techniques, and transfer to other beamlines. The present article begins with an explanation of soft x-ray ptychography with SOPHIE (Sec. II), followed by a description of the SOPHIE chamber (Sec. III) and an example of high resolution ptychographic imaging to demonstrate the performance of SOPHIE (Sec. IV).

## II. PRINCIPLES OF PTYCHOGRAPHIC IMAGING WITH SOPHIE

The standard configuration of SOPHIE is defocused probe ptychography,[64] a sketch of which is depicted in Fig. 1(a). The permissible movement directions of the components are also shown in Fig. 1(a), with $z$ defined as the x-ray beam direction. The incident monochromatic x-ray beam is shaped by a FZP onto the sample. In contrast to STXM, x-ray ptychography does not require the sample to be in focus of the x-ray beam in order to maximize the spatial resolution. Rather, the FZP is used for flexible beam shaping, which leads to several advantages. First, the high divergence of the beam means that the bright-field cone of the beam is spread over a large area of the detector, reducing the requirements for high dynamic range and removing the need for a beamstop in front of the detector. Second, the FZP introduces phase structure to the probe, providing greater stability to the phase retrieval process.[65] Third, the controllable positioning of the sample away from the focal plane of the FZP allows the size of the sample illumination to be tailored and adjusted to the scan area. The overlap in real space, which is critical for defocused probe ptychography, is determined by the ratio between probe diameter and the step size of the scan. Defocused probe ptychography offers another benefit for the imaging of radiation sensitive samples by reduction of the radiation dose rate per area due to the larger x-ray beam.

The undiffracted central part of the x-ray beam passes through the FZP as zero-order radiation with an unmodified shape and it must be blocked in order to avoid overpowering the flux of the focused beam and saturation of the detector. This is accomplished by a central stop (CS) and the order-selecting/sorting aperture (OSA). The CS is a disk of a few micrometer thickness, usually made of Au, which is opaque to the soft x-rays. Some FZPs include a CS in their middle on the silicon nitride ($Si_3N_4$) membrane. However, certain FZPs rely on a separate freestanding CS that is positioned in front of the FZP.

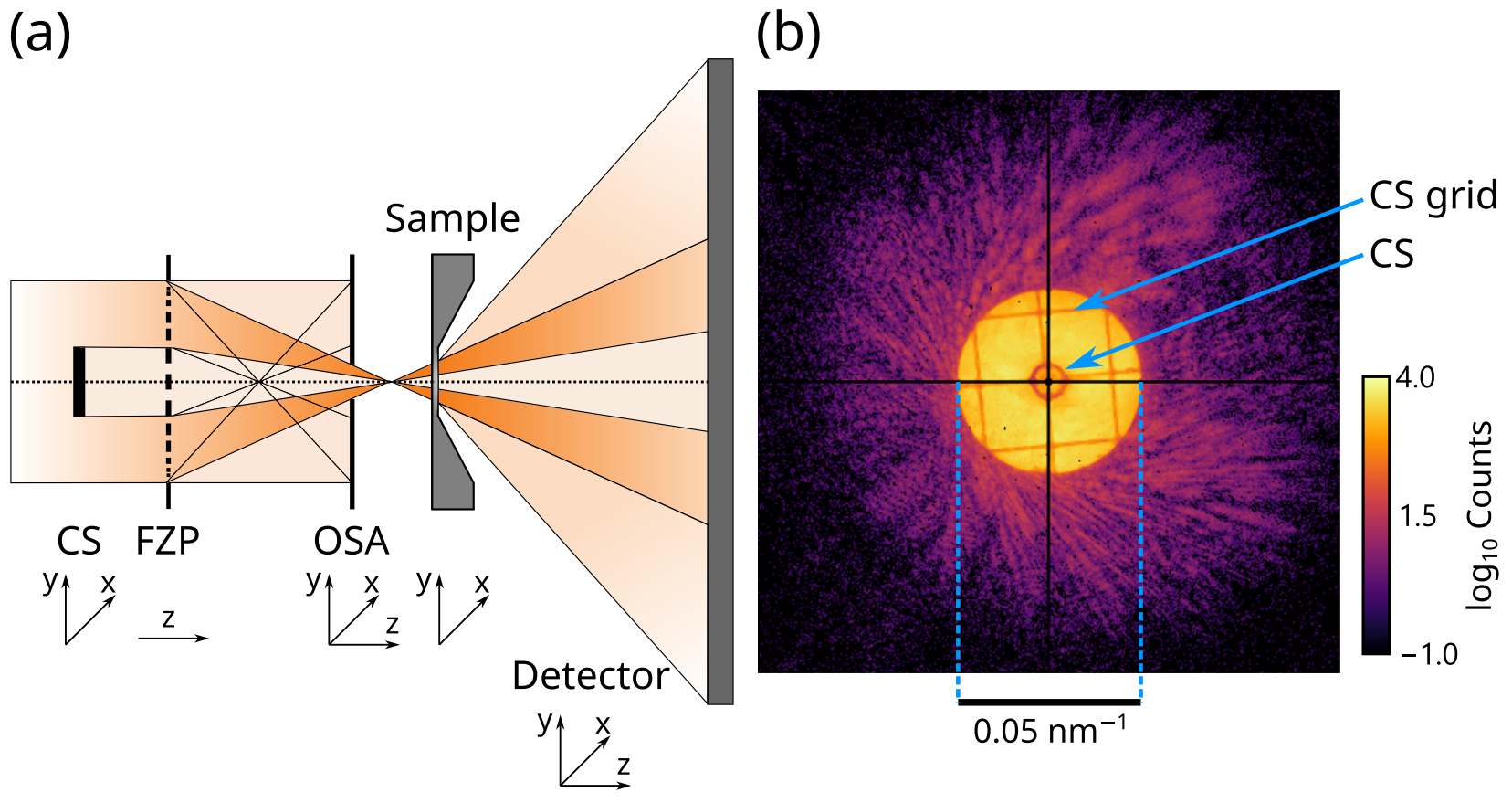

**FIG. 1.** (a) Sketch of the ptychography setup in the defocused probe mode employed in SOPHIE using a coherent monochromatic soft x-ray beam (not to scale). The size of the x-ray spot on the sample during a ptychographic scan is selected by moving the FZP in z-direction. The permitted movement directions of the respective components in SOPHIE are indicated by the arrows. Higher order foci of the FZP, one of which is indicated by the broken lines, are blocked by the OSA. The CS and OSA remove the undiffracted zero-order radiation. (b) Detector image of a microdiffraction pattern, a set of which constitutes ptychography data. The annulus is formed by the far-field diffraction of the first-order focus of the FZP and the CS, which is fixed at the center of a grid. The diameter of the central cone, which constitutes the bright-field image, is reciprocal to the outer zone width of the FZP. The outer-zone width is 20 nm for the displayed image. The dark-field information is contained in the scattered photons outside the central cone. The black cross through the center of the image is due to a gap between four individual detector chips.

The OSA is a small circular aperture cut through a metal sheet that is placed upstream of the sample. A FZP is a diffractive optic that has several diffraction orders with distinct foci and their removal by the OSA is imperative for STXM or ptychography measurements. The even-order foci are suppressed by FZPs that have equal widths of transparent and opaque rings, which means that only the odd-numbered foci must be obstructed.[17,18] The OSA commonly has diameter below 100 μm and must be smaller than the CS to shield the sample from remaining zero-order radiation. The permissible OSA-sample distance for which the OSA does not block the focused x-ray beam depends on the FZP, but is usually in the order of a few hundred micrometers. This is a challenge compared to hard x-rays, where focal distances are longer. Furthermore, the OSA also ensures high spectral purity by blocking higher-order x rays that may still be present in the beam and are also focused by the FZP. The removal of the zero-order radiation by the combination of CS and OSA avoids the usage of a beam stop that is placed directly in front of the pixelated area detector in other CDI experiments, which leads to the loss of low-frequency data and decreased quality of the reconstructed image.

Having thus shaped the x-ray beam, it is then scattered by the sample and the diffraction patterns are recorded by a pixelated area detector. An example of a microdiffraction pattern from a ptychographic dataset is shown in Fig. 1(b). The data shown here were recorded with a single-photon counting inverse-Low Gain Avalanche Diode (iLGAD) EIGER detector that consists of four iLGAD Eiger chips with a gap of 150 μm that causes a radiation insensitive cross that quadrisects the image.[49] The far-field diffraction pattern of a single focus from the FZP forms a cone of high intensity that appears as an annulus on an area detector. The symmetric annulus only appears when the OSA is correctly aligned with respect to the center of the FZP and is suitably close to the focal plane. While the point detector of STXM supplies only minimal information so that each optical element must be scanned across the beam in order to align the instrument, the two-dimensional detector of ptychography provides direct feedback on the positioning of the optical components for swift alignment. A misaligned OSA cuts the focused x-ray beam, which becomes apparent as an asymmetry or disappearance of the annulus—note that the image is flipped at the focal plane because the OSA is upstream of the focus. Complete removal of the OSA from the x-ray beam path drastically increases the radiation on the detector, which necessitates attenuation of the x-ray beam, for instance, by detuning the undulator, to avoid damage of the pixelated area detector due to overexposure.

The selection of the illumination spot diameter on the sample typically relies on placing the sample in focus to have a fixed reference from which to move the FZP. This is achieved by locating a high contrast feature and performing a knife-edge focus scan as in STXM, whereby the high contrast feature is scanned through an x-ray beam over a range of different FZP distances and integrating the transmitted flux on the detector. The focus is identified as the FZP-sample distance that resulted in the sharpest transition between maximum and minimum transmission of the beam. Alternatively, the pixelated area detector also enables setting of the focus by analysis of the shadow image of the sample, which flips in the central cone when the focal plane is crossed.

As the focal lengths of FZPs in the soft x-ray range are in the order of several millimeters, a displacement of ~10 μm from the focal plane leads to a widening of the x-ray beam to a spot size of around 1 μm. The precise value of the illumination size at a given distance of the sample to the focal plane is calculated from the specifications of the FZP.

In the diffraction pattern shown in Fig. 1(b), the CS is separate from the FZP and in the center of a rectangular grid. The sample also contributes to the central cone and this bright-field is the dominant contribution to STXM images. Hence, integration of the microdiffraction patterns yields a STXM image that can be used for navigation. As defocused-probe ptychography is usually carried out with the sample slightly downstream of the focus, an in-line Gabor hologram is present.[66] The sample may also be moved far from the focal plane to obtain a demagnified x-ray shadow microscope image,[67] which may be utilized for coarse navigation on the sample. Apart from this, an inhomogeneous illumination of the FZP can also be gauged by inspection of the central cone. This can be corrected by realignment of the beam or the endstation itself. Outside of the central cone are scattered photons that form the dark-field data, which can be several orders of magnitude less intense than the radiation in the cone.[68,69] High spatial resolution requires accurate retrieval of the phase from high angles on the detector.

A ptychographic scan consists of the acquisition of a set of coherent diffraction patterns and their reconstruction in a complex-valued image of the sample and illumination. Four requirements must be fulfilled in order for ptychography to yield high resolution images.

1. Sample positioning: Several diffraction patterns must be obtained from known positions of the sample. The sample must be translatable in the plane normal to the incident x rays with nanometric precision. Mechanical instabilities smear out the diffraction pattern to the detriment of the spatial resolution. In SOPHIE, the increased sensitivity to motion is combated via differential interferometry between the FZP and sample, which is explained in Sec. III D. Ptychography also relies on the redundancy of information from diffraction patterns with overlapping areas. The degree of overlap can be modified by variation of the FZP–sample distance and the step size in the scan.
2. Coherent illumination: Ptychographic imaging requires an illuminating wavefront of as high coherence as practical. Synchrotron radiation emitted from undulators in the soft x-ray regime offers high coherent flux and control of the x-ray polarization.[70] The coherent component of the flux is much higher with soft x rays as it scales with the square of the wavelength ($\lambda^2$) and most soft x-ray undulator beamlines at third generation light sources are already diffraction-limited.[71] The upgrade to fourth-generation synchrotron radiation facilities promises further improvements of flux and coherence in the hard x-ray regime, but will not affect soft x-ray ptychography as substantially.

   Laboratory sources in the soft x-ray range do not currently exist with sufficient coherence for ptychography, although it is possible in the extreme ultraviolet radiation (XUV) energy range with the coherent photon flux from high-harmonic generation (HHG) sources both in transmission[72–75] and in reflection.[76]
3. Detection of photons: The detector benefits from low noise and high sensitivity to record faint x-ray scattering from small features at high angles for phase retrieval, as well as high dynamic range to simultaneously detect the low-frequency information from the central cone. The majority of ptychographic imaging in the soft x-ray regime in chambers other than SOPHIE was performed with CCD[28–35,50–63] and scientific complementary metal–oxide–semiconductor (sCMOS)[36,38–40,42–45] detectors. Single-photon counting area detectors with no readout noise and high dynamic range are ideal for ptychography. Such detectors have existed in the hard x-ray regime since the 2000s[77] and have become the mainstay for ptychography in that energy range. The availability of such detectors for soft x-rays is more recent.[49] The stringent requirements that ptychography places on photon detection may have contributed to the more rapid development of hard X-ray ptychography compared to its low energy counterpart.
4. Reconstruction algorithm: Ptychographic datasets are input to iteratively compute complex-valued images of the sample with dedicated reconstruction algorithms. These extract both the object and probe (illumination) functions from the diffraction patterns simultaneously.[78–80] The probe can be iteratively refined from an initial model probe function, which is usually undertaken with the dataset of a strongly scattering sample. The resulting probe function can then be used to initiate the reconstructions for the remaining measurement campaign. Ptychographic algorithms were first developed for hard x-ray ptychography and can be directly applied to datasets obtained with soft x-rays. The reconstruction process is vastly accelerated by parallelized computation on graphics processing unit (GPU) clusters. The most popular algorithms have proven to be difference-map[78,81] and extended ptychographical iterative engine (ePIE),[80,82] the results of which can be further refined with the maximum-likelihood method.[83,84] Moreover, reconstruction algorithms can account for partial coherence of the illuminating wavefront by decomposition of the probe function into several modes.[85] This is also key to the reconstruction of ptychographic data obtained in continuous "fly-scan" mode.[86–88] In addition, orthogonal probe relaxation was developed to address an illumination that changes during the ptychographic scan.[89] Prominent examples for Open Access ptychographic reconstruction programmes are PtyPy[90] and PtychoShelves,[91] which are written in Python and MATLAB, respectively. The reconstructions are performed after the completion of a ptychographic scan (offline), but real-time reconstructions may play a role in expediting analysis in the future.[28,92,93]

In summary, the resolution limit in ptychography is defined by the accurate recovery of the phase from high scattering angles, coherent flux, wavelength of the x-rays, and the stability of the sample positioning. Essentially, only the last of these is related to the instrumentation of the endstation, with the exception of the detector. The following sections provide insight into the optical components within SOPHIE and how they ensure ptychographic imaging with high spatial resolution.

## III. THE SOPHIE MICROSCOPE

The exterior and interior of the SOPHIE microscope are displayed as computer-aided design (CAD) drawings in Figs. 2 and 4, respectively. Furthermore, Fig. 3 contains photographs of both. Soft x-ray ptychography with SOPHIE complements the STXM at the PolLux beamline,[94] hard X-ray ptychography at the cSAXS beamline, and the Electron Microscopy Facility (EMF) within the Paul Scherrer Institute for studies at different length scales.[95]

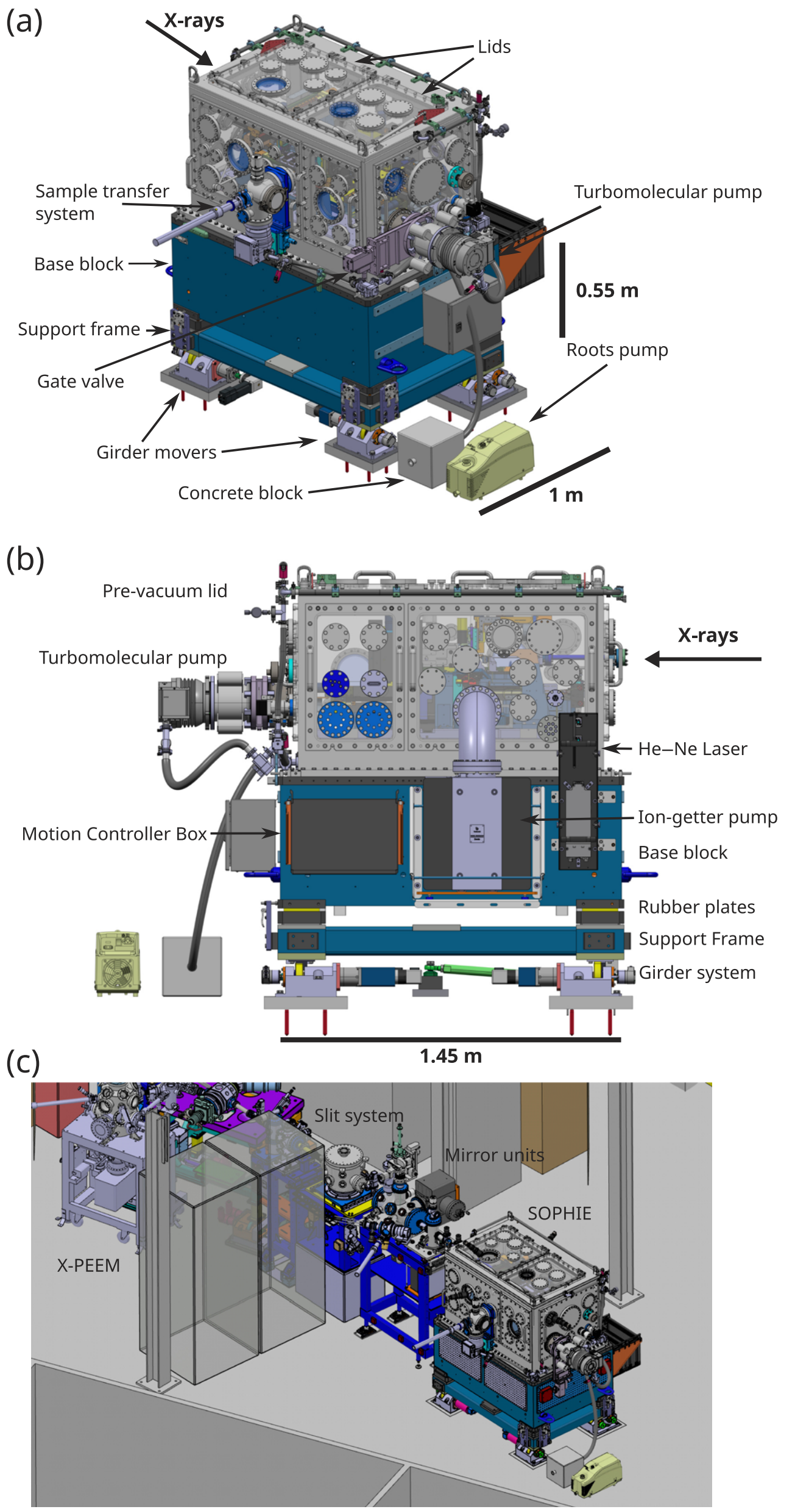

**FIG. 2.** (a)–(b) Exterior of the SOPHIE microscope. (c) SOPHIE at the SIM beamline at the SLS 2.0.

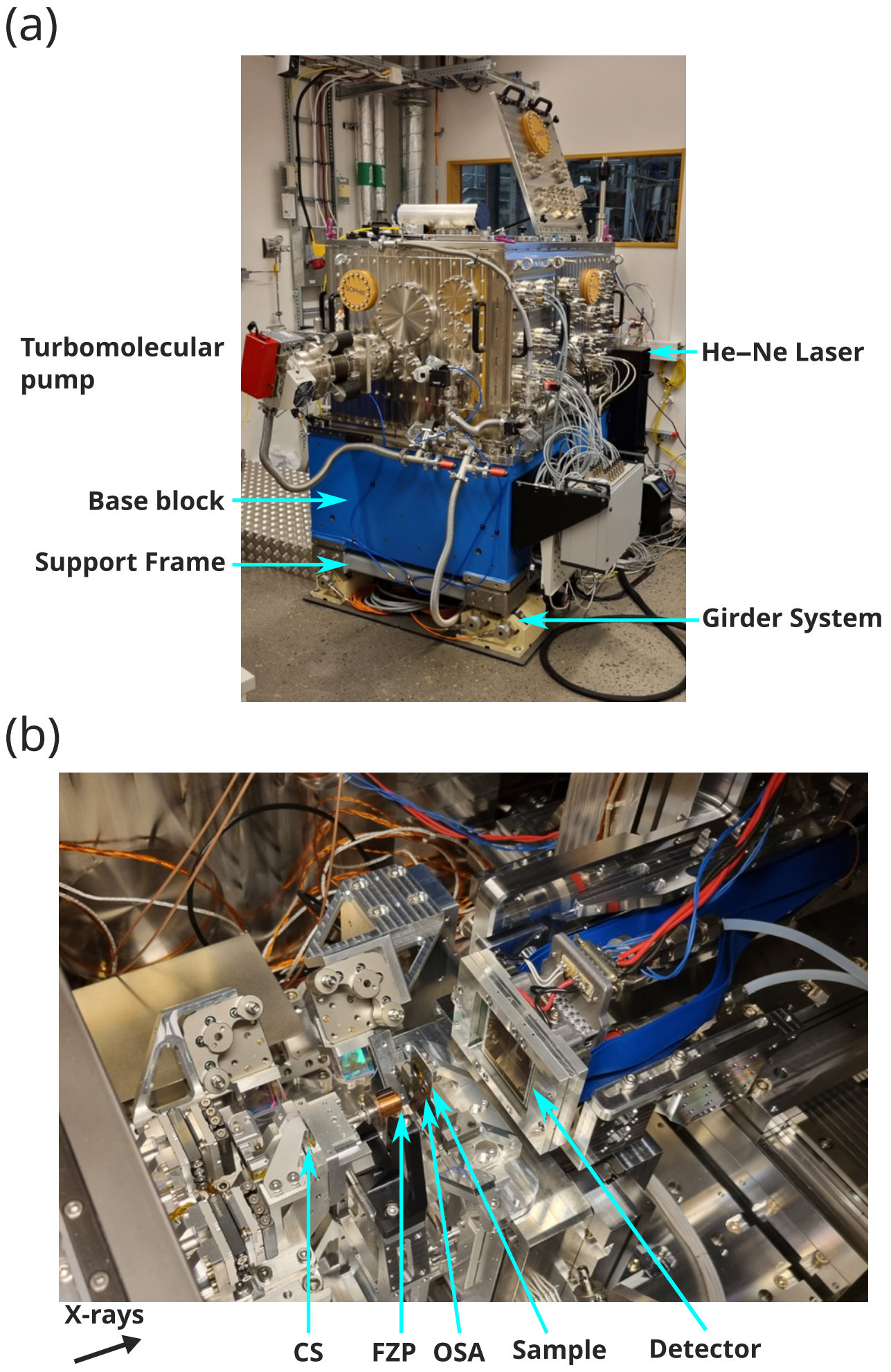

**FIG. 3.** (a) SOPHIE microscope at the SoftiMAX beamline in January 2024. (b) Interior of the SOPHIE chamber with an iLGAD EIGER detector for soft x-ray ptychography; see Fig. 4(c) for a schematic side view.

### A. Beamline and peripheries

The SOPHIE chamber is based on the SIM beamline of SLS 2.0 at the Paul Scherrer Institute[96] [see Fig. 2(c)] and was located at the SoftiMAX beamline of MAX IV (Lund, Sweden) from January 2024 to June 2025 during the SLS 2.0 upgrade shutdown [see Fig. 3(a)]. Both soft x-ray beamlines have elliptically polarizing undulators with an APPLE II at SoftiMAX and an APPLE X with knot design[97] at SIM following the SLS 2.0 upgrade. At the SIM beamline, the soft x-ray beam is steered through the x-ray photoemission electron microscope (X-PEEM) that is permanently at the beamline and after which SOPHIE is placed [see Fig. 2(c)]. The spatial coherence of the x-ray beam is ensured with an adjustable slit-system 3 m away from the sample that acts as a spatial filter, defining the secondary source.

The vacuum chamber of the endstation is mounted on a 2.4 t base block ($1.45 \times 1.00 \times 0.55$ m$^3$) that is placed on top of a support frame above a girder mover system at the beamline. The total weight of SOPHIE is 4.3 t and the entire endstation can be hoisted by a crane, which allows exchange with other endstations at the open port of the SIM beamline or the transfer of SOPHIE to various beamlines with potentially different energy ranges.

Translation and rotation of the entire endstation with respect to the x-ray beam is enabled reproducibly at micrometer precision via the girder movers.[94] This allows translations in two directions (x: horizontal and y: vertical) and rotations in three directions (yaw, pitch, and roll). Fine lateral alignment of the FZP with respect to the x-ray beam is carried out with the girder movers.

The endstation is only fixed to the base block and not the girder mover support frame. Instead, it is decoupled by rubber plates (isoloc Schwingungstechnik GmbH, IPL 10) to ensure dampening of vibrations from the ground.

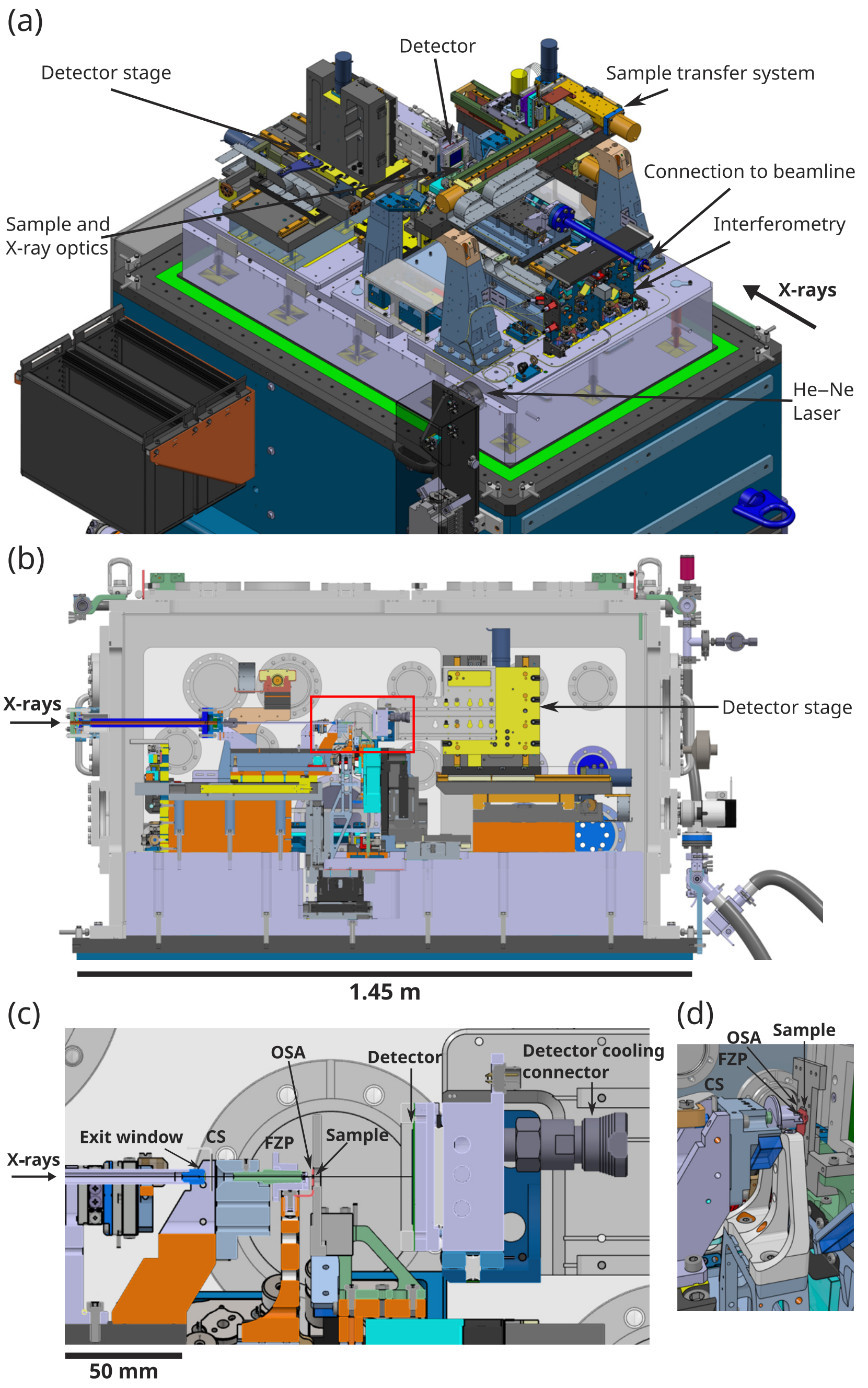

**FIG. 4.** (a) Interior of the SOPHIE chamber. (b) Side view of interior parallel to the x-ray beam. (c) and (d) Magnified views of the soft x-ray ptychography setup framed in red in panel (b). The schematic x-ray beam path with movement directions of the components is shown in Fig. 1(a). The FZP-sample and sample-detector distances are typically in the order of 5 and 100 mm, respectively. The gap between the OSA and sample is below 1 mm; see Fig. 3(b) for a photograph of the soft x-ray ptychography setup.

## B. SOPHIE chamber

A base pressure of $5 \times 10^{-7}$ mbar is achievable by a turbomolecular pump (Pfeiffer Vacuum Technology AG) with the additional option of an ion-getter pump for improved vacuum. A roots pump provides the rough vacuum and is connected to the turbomolecular pump by bellows that pass through a concrete block, in order to minimize vibrations (see Fig. 2).

The interior of the vacuum chamber can be accessed by two lids on the top [see Fig. 2(a)] and several viewports allow monitoring of the interior while the chamber is under vacuum. Samples can be changed manually; however, SOPHIE is also designed to function with an automated sample transfer system operated with a load-lock that avoids venting of the chamber.

A gate valve isolates the turbomolecular pump from the vacuum chamber while venting, which allows it to remain in operation at full rotational speed. The gate valve is opened once rough vacuum is re-established in the chamber. A full vent and pumping cycle takes at least 45 minutes with additional time necessary for warming up and cooling down the detector.

Upon connection of the chamber to the beamline and achievement of a sufficiently high vacuum ($10^{-5}$ mbar), soft x-rays enter the chamber through a flight tube in the form of an edge-welded bellow [see Figs. 4(a) and 4(b)]. This leads to a window of $Si_3N_4$ (50 nm thick and $500 \times 500$ $\mu m^2$ in area) that separates SOPHIE from the beamline vacuum and can permit operation at lower vacuum levels than otherwise allowed.

The emerging x-rays are shaped by the x-ray optics [see Fig. 4(c)], details of which are discussed in Sec. III C. The x-rays from the FZP impinge on the sample that is movable on a piezoelectric stage, which is described in Sec. III D. The transmitted and scattered x-rays are recorded with a soft x-ray detector (Sec. III E).

All of the x-ray optics, the sample, and the detector stages are motorized. The directions of movement are shown in Fig. 1(a). SOPHIE is controlled with Pixelator STXM Control Software that was first used at the PolLux beamline[94] and is now employed at several STXM beamlines in Europe. All scan data are recorded in NeXus compliant HDF5 files.[98]

Three-dimensional imaging with ptychographic soft x-ray laminography can be performed with SOPHIE. This approach is particularly well-suited for soft x-ray microscopy due to the reduced effective thickness that avoids the missing wedge artifact caused by high absorption of flat samples under high angles in standard tomographic imaging. Previous experiments have been successfully performed with STXM[99–101] and hard x-ray ptychography.[102,103] A rotatable sample stage must be installed at an appropriate tilt angle in order to provide the required sample projection geometries in laminography. In the case of SOPHIE, the same custom made stage (Smaract GmbH) used for STXM laminography at the PolLux beamline can be installed at 45° with respect to the incident beam.[99]

Static magnetic fields can be applied to the sample by means of a dipole magnet in both in-plane and out-of-plane configuration with maximum values of 150 and 210 mT, respectively. This magnet setup cannot be seen in Fig. 3(b) and Fig. 4, as it is not installed for standard measurements. The magnetic field setup is currently restricted to two-dimensional imaging.

A 4 K cryo link to a liquid He cryostat (LT3B Helium Flow Cryostat) can be mounted for imaging at cryogenic temperatures. This allows cooling of the sample to below 20 K. A customized sample and OSA stage with polyetheretherketone (PEEK) isolation are in place for this. Cryogenic temperatures are highly relevant for the study of samples that undergo phase transition in condensed matter physics and for mitigating radiation damage in the study of biological samples without the need for chemical fixation.[104] The automated sample transfer system with the load-lock is mandatory for the majority of such experiments.

## C. X-ray optics

A view of the x-ray optics setup in the SOPHIE chamber is shown in Figs. 3(b), 4(c), 4(d), and 5. The movement directions of the individual components are indicated by the arrows shown in Fig. 1(a). The coordinate system is defined as $z$: x-ray beam direction ($+z$: away from the source), $x$: horizontal ($+x$: outboard from the ring), and $y$: vertical ($+y$: up) directions.

The beamline exit window on the edge-welded bellow is motorized (SLC-2430 SmarAct GmbH). It can be centered on the beam in horizontal and vertical directions with travel ranges of 2 mm, which is restricted by the bending of the bellow. The exit window can be used to shape the incident beam for the illumination and selection of an individual FZP from an array of several closely spaced FZPs on a $Si_3N_4$ membrane.

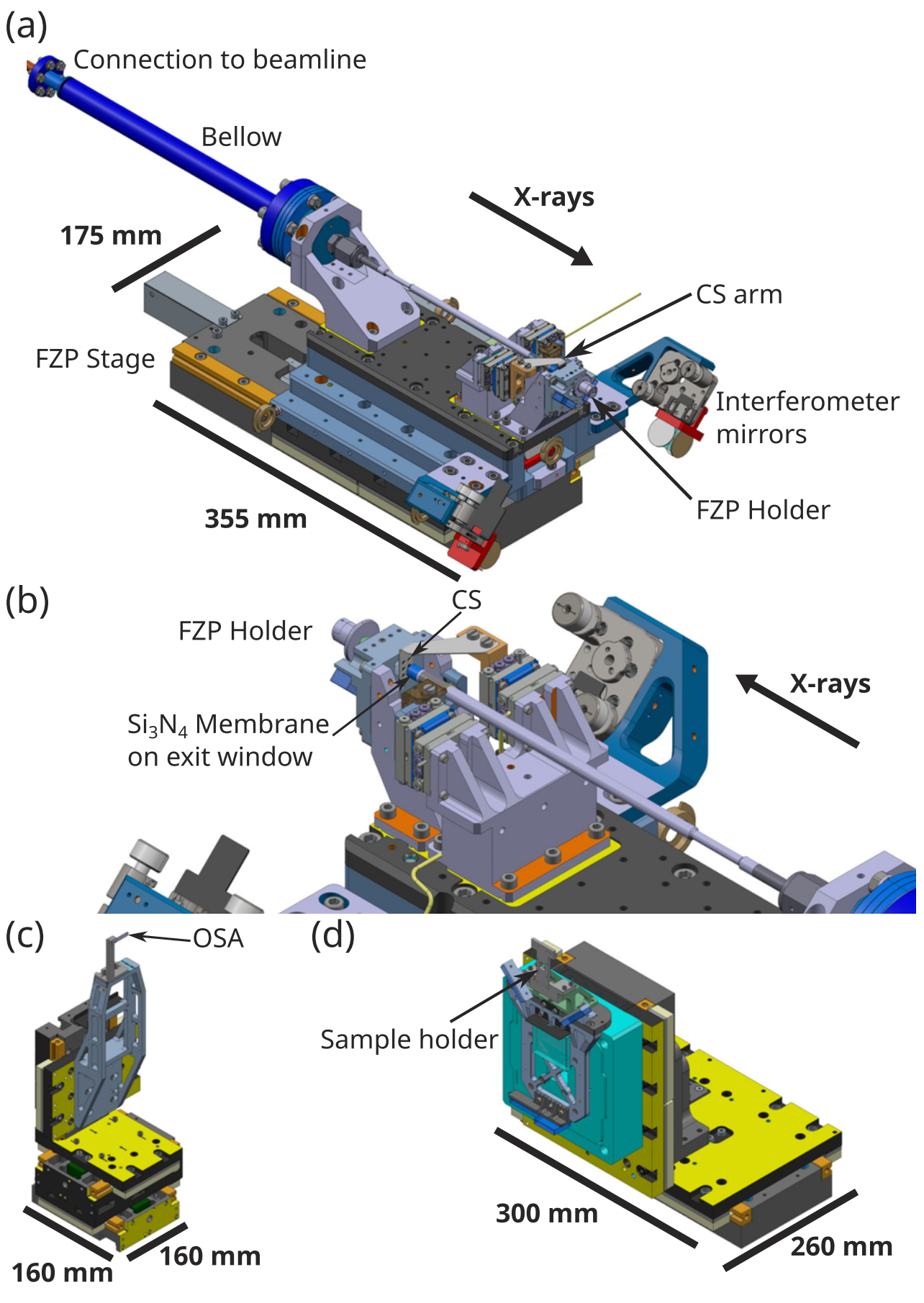

**FIG. 5.** Stages and components of the x-ray optics in the SOPHIE chamber. See Fig. 1(a) for schematic view and movement directions. Panels (a) and (b): FZP stage with exit window and CS. Panel (c): OSA stage. Panel (d): sample Stage. Both OSA and sample stage are sunk into the base block for stability [see Fig. 4(b)].

In order to align the FZP with respect to the center of the x-ray beam, the girder movers are used to move the entire chamber in horizontal and vertical directions. The FZP itself is only movable upstream and downstream along the propagation direction of the x-rays by a PI N-331 PICMAWalk piezoelectric walking drive with a range of 150 mm. The entire stage is shown in Figs. 5(a) and 5(b). The FZP–sample distance is dependent on the focal length of the chosen FZP and is energy-dependent. Usual values are in the order of 5 mm.

A separate CS is available on a stage in front of the FZP [see Figs. 5(a) and 5(b)]. The CS location is controlled by two linear piezo positioners (SLC-2430 SmarAct GmbH) with a travel range of 16 mm laterally to the x-ray beam. The standard CS is circular with a diameter of ~100 μm and a thickness of several micrometers of Au. The CS is fixed on a rectangular grid [see Fig. 1(b)] and glued to the motorized holder.

The OSA in SOPHIE can be moved in three dimensions by 20 mm. The stage is shown in Fig. 5(c) and is motorized by a PI N-331 PICMAWalk walking drive. A large part of the stage is sunk into the base block [see Fig. 4(b)] to minimize vibrations. The spacing between the OSA and sample is less than 1 mm, but the exact value is energy-dependent. Thus, the OSA position must be adjusted after significant changes in photon energy.

## D. Positioning and motion control

Samples for soft x-ray transmission measurements are usually prepared on $Si_3N_4$ membranes with Si frames or Cu transmission electron microscope (TEM) grids. These can be attached to different sample holders that are meant for standard ptychography, laminography, or cryogenic cooling. Coarse movement of the sample stage (Steinmeyer Mechatronics GmbH) is possible in horizontal (PI N-216 PICMAWalk Walking Drive) and vertical (PI N-331 NEXLINE Linear Actuator) directions with ranges of 20 and 10 mm, respectively. While there is no coarse sample movement in the direction of the x-ray beam, Pixelator control software is configured with a virtual stage that passes move requests to the FZP and OSA stages in order to mimic z-axis motion of the sample. The removal of the sample z-motion is beneficial for the integration of different sample environments, such as the connection to the cryostat. In addition to this, the sample–detector distance is kept constant without the requirement for a large linear stage to move both the sample and the detector assembly along the z-axis. Movement along the z-axis is only possible with the detector (see Sec. III E).

For ptychographic measurements, the sample must be scanned across the beam with nanoscale accuracy. This is achieved by means of a piezoelectric scanner with a range of 100 μm and six degrees of freedom (nanoFaktur GmbH, nPoint, Inc., custom-made). The coarse stage is engaged as soon as the range of the fine stage is exceeded. Although the weight of the SOPHIE microscope moves the resonance frequencies of the chamber outside of the typical range of environmental noise sources, the presence of mechanical vibrations cannot be completely suppressed. In order to compensate for these, the sample position is interferometrically measured relative to the FZP and active feedback is applied to the piezoelectric stage for correction of the sample position. In addition to the closed-loop feedback, operating on a real-time Linux system, the interferometer positions are acquired using a PandABox (Position and Acquisition Box, Quantum Detectors Ltd.), which enables synchronous recording with high-speed data transfer. The default sample scanning is performed step-wise to ensure settling of the sample for high spatial resolution, although continuous scanning trajectories[86–88] can also be chosen as operation modes.

A frequency-stabilized He–Ne laser (Zygo Corporation) is attached to the base block and the shielded laser light enters the chamber through an optical viewport [see Fig. 4(a)]. The laser light is split into beams that feed a set of heterodyne, double-pass interferometers[105] that measure the distance between FZP and sample in three dimensions, as shown in Fig. 6. In addition, rotational error motions of the sample stage are measured by a separate set of interferometers (not shown). Figure 6(a) shows a schematic of the z-axis interferometer that is sensitive to the path length between the fixed reference attached to the FZP stage and a flat mirror attached to the sample stage. This interferometer is most relevant for three dimensional imaging with laminography and for STXM imaging with high resolution FZPs that have a short depth of focus. A schematic of the interferometers that measure the relative lateral position of flat mirrors on the sample and FZP stages is displayed in Fig. 6(b). These are inclined by ±45° with respect to the vertical direction. The directions are denoted as j and k-axes in Fig. 6. The oblique position measurements are converted to horizontal and vertical positions in the software for the use of conventional x and y-axes. Each of these interferometers includes a corner cube optic that provides an offset between the input and output beams, as well as assisting alignment of the optics. The arrangement of the measurement arms of the interferometers is depicted in Fig. 6(c).

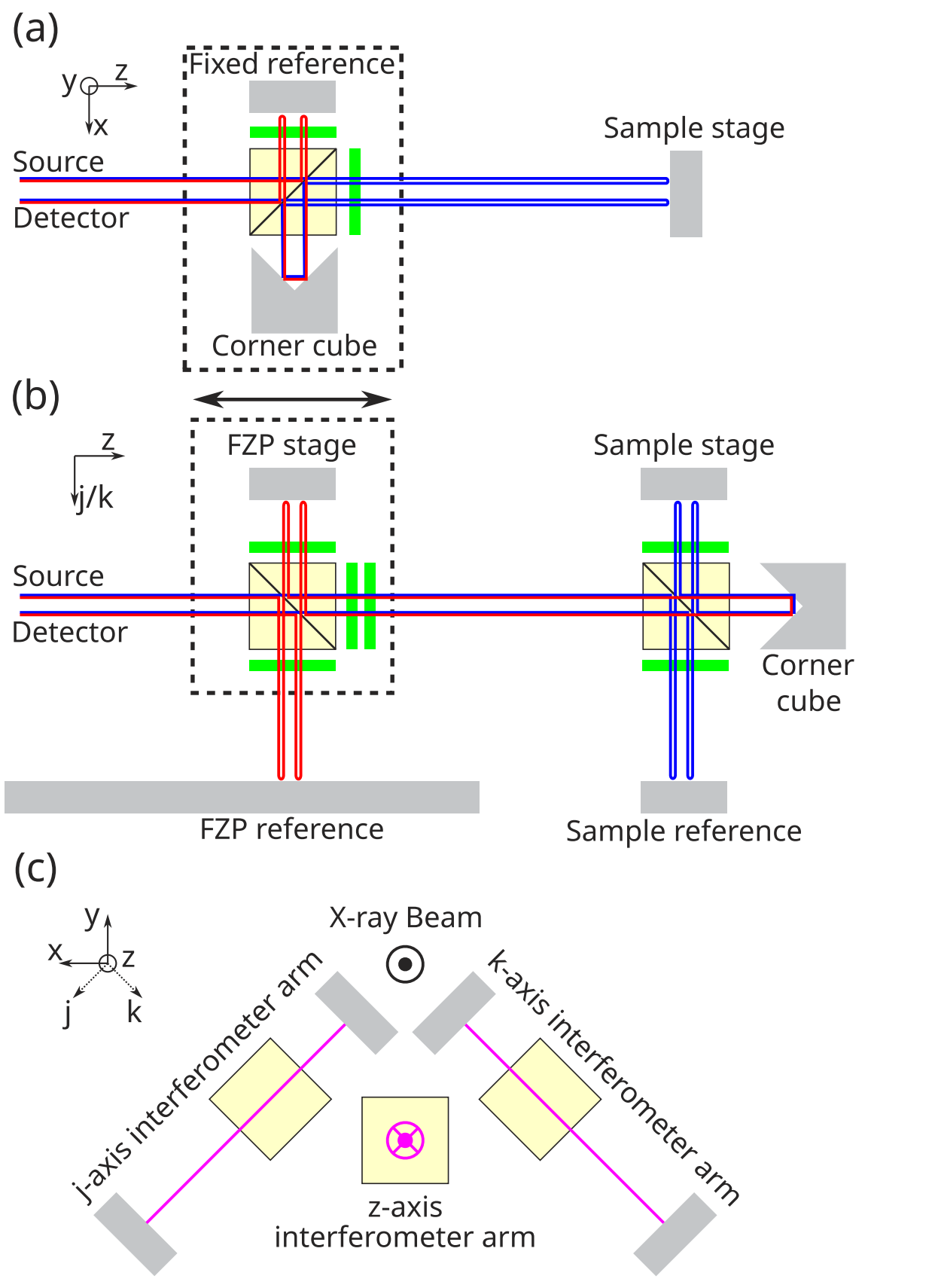

**FIG. 6.** Interferometric setup for FZP-sample position measurement. The red and blue lines symbolize the two orthogonal linear polarizations of the laser beam. (a) Top view sketch of the heterodyne double-pass interferometer that measures in the x-ray beam direction (z axis). (b) Orthogonal view sketch of the heterodyne double-pass interferometers measuring lateral to the x-ray beam at ±45° to the vertical (j/k axes). The green objects are quarter-wave plates and the dashed-line boxes indicate the group of objects that are mounted on the FZP z-stage and move together, which is pointed out by the arrow. (c) Sketch of the arrangement of the interferometer measurement arms in the SOPHIE chamber.

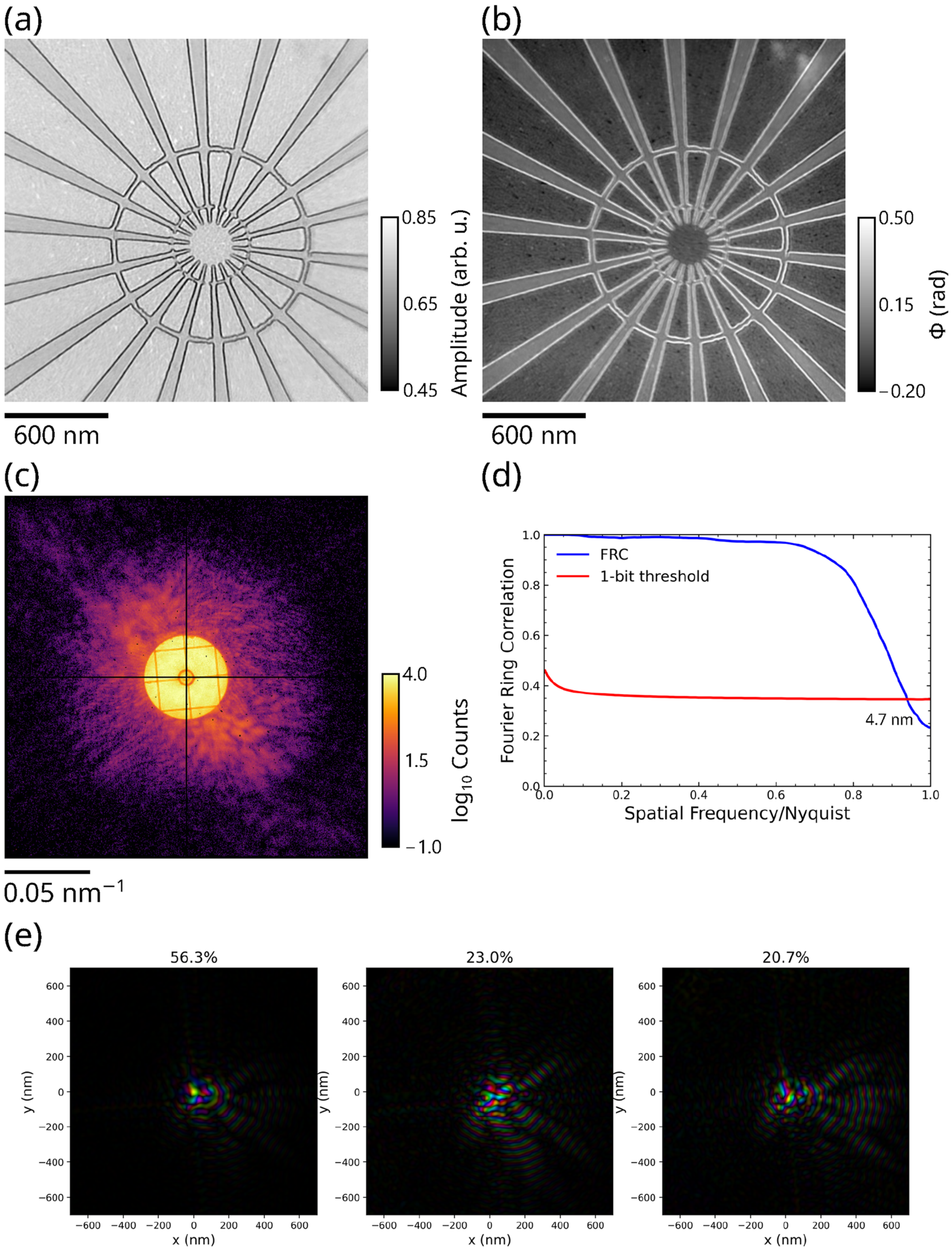

FIG. 7. Ptychographic imaging of a Siemens star at 706 eV. (a) Amplitude image. (b) Phase image. (c) Microdiffraction pattern from the raw ptychographic dataset that shows x-ray scattering at high angles. The low noise and high dynamic range are a hallmark of the single-photon counting iLGAD EIGER detector.[49] (d) A spatial resolution of 4.7 nm is indicated by the FRC with the 1-bit criterion. (e) The reconstructed probe from the FZP (line doubled Ir 500 μm diameter, 20 nm outer zone width, randomized zone variation) was decomposed into three modes, the respective weights of which are shown above in percent.

In order to cover a wide photon energy range, large translations of the FZP in z-direction are required. The design of the interferometer optical layout permits such movements up to 170 mm without the need of installing long mirrors on moving components. The 45° tilt of the interferometer axes permits rigid mounting of the reference mirrors close to the base block and frees additional space is left open to the sides of the sample that can be used for any additional apparatus that modifies the sample environment (e.g., magnets and gas cell).

## E. Detectors

The interior of the SOPHIE microscope is spacious and offers flexibility concerning the installed detector. The detector is bolted onto a stage that can be moved with stepper motors (Phytron GmbH) in three directions with travel ranges of 200 mm for horizontal direction and along the optical axis, as well as 50 mm in vertical direction. This stage is designed to sustain a weight of up to 10 kg and forces up to 150 N so that it is compatible with a bellow that may be used to connect to a detector for cooling and interface feedthrough to vacuum. Several flanges allow for different connections for cooling, data transfer, and power to the detector.

The adaptability of the sample–detector distance permits soft x-ray ptychography at a wide range of photon energies as the sampled reciprocal space changes. In principle, measurements in the tender x-ray range above 2000 eV are achievable with higher sample–detector distances. Furthermore, it is possible to switch to a simultaneously mounted avalanche photodiode for STXM without venting the chamber and replacement of the pixelated area detector. This benefits time-resolved studies that require the high temporal resolution that an avalanche photodiode can offer.[47]

The standard detector for the SOPHIE microscope in the beginning of its operation has been the iLGAD EIGER, which is currently the only single-photon counting soft x-ray detector.[49] The best performance is achieved for energies above 600 eV, although its functionality has been demonstrated down to the O K-edge at 530 eV.[48] Hitherto, it has been successfully employed for the imaging of multiferroic domains, non-collinear magnetism, and magnetic domain wall dynamics.[46–48] The iLGAD EIGER has a pixel size of 75 μm and the sensor consists of four 256 × 256 pixel chips that form an area of ~40 × 40 mm$^2$. The standard sample–detector distance for operation with this detector is around 100 mm.

The flexible design of SOPHIE allows for other detectors such as the AXIS-SXR soft X-ray sCMOS camera (AXIS Photonique, Inc.) with a smaller pixel size of 11 μm and fewer constraints at lower photon energies.[106,107]

## IV. INSTRUMENT PERFORMANCE

SOPHIE was commissioned at the SIM beamline of the SLS and the SoftiMAX beamline at MAX IV. Figure 7 shows the result of a ptychographic scan of a Siemens star at 706 eV (1.76 nm) with the iLGAD EIGER detector.[49] The fabrication method of the Ir Siemens star is described in Ref. 108. The reconstructed amplitude and phase are depicted in Figs. 7(a) and 7(b), together with an exemplary raw diffraction pattern in Fig. 7(c). An exposure time of 200 ms was selected at a sample–detector distance of 96 mm. A line doubled Ir FZP with 500 μm diameter, 20 nm outer zone width, and randomized zone variation was used to create a 400 nm FWHM illumination spot on the sample. The scan points followed a Fermat spiral[109] of 50 nm step size with a dwell time of 220 ms at each position, which included a 20 ms wait time for settling of the sample at each position. This comparatively long dwell time was chosen in order to maximize the spatial resolution by obtaining the highest quality diffraction patterns possible.

The PtychoShelves software package[91] was employed for the reconstruction of the ptychographic dataset. This was performed with 1000 iterations of difference-map[78,81] and 300 iterations of maximum-likelihood refinement[83,84] with three probe modes.[85] It took ~1 min to reconstruct this dataset. The reconstructed probe modes are shown in Fig. 7(e). The probe modes are anisotropic due to the randomized zone variation of the FZP that imparts phase structure to the probe.[110] The pixel size of the reconstructed images is 4.4 nm.

The sample is a strong x-ray scatterer, with photons detected until the edge of the detector [see Fig. 7(c)]. The reconstructions show high spatial resolution and are free of reconstruction artifacts. A quantification of the imaging reproducibility is possible by estimation with a Fourier Ring Correlation (FRC) between two identical ptychographic scans,[111,112] which is plotted in Fig. 7(d). The 1-bit threshold indicates a half-pitch spatial resolution of 4.7 nm, which is close to the limit imposed by the reconstructed pixel size.

It is to be noted that the maximum achievable spatial resolution is sample-dependent. Objects with small features and intense x-ray scattering can be routinely imaged with sub-10 nm spatial resolutions. However, ptychography can be prone to artifacts when samples have large featureless areas without scattering that typically appear as an imprint of the probe at the scanning positions in the reconstructed image. Exposure times vary from sample to sample similarly to those in STXM measurements. Shorter exposure times are possible with a maximum continuous frame rate of 2 kHz in 12-bit mode for the iLGAD EIGER detector.[49]

## V. SUMMARY AND OUTLOOK

The SOPHIE microscope accommodates a broad range of microspectroscopic experiments with soft x-ray ptychography. High spatial resolution is ensured by a single-photon counting detector and interferometric control of the sample positions. The APPLE X undulator at the SIM beamline of SLS 2.0 promises high coherent flux with control of the polarization of the soft x-rays and high spectral resolution, which is key for experiments in condensed matter physics, chemistry, and environmental and life sciences. Three-dimensional imaging at high spatial resolution will be enabled by soft x-ray ptychographic laminography, which has previously been achieved by STXM at the POLLUX beamline.[99–101]

Further improvements in detector technology are expected to result in single-photon counting detectors that can access the water window between the K-edges of C (280 eV; 4.4 nm) and O (530 eV; 2.3 nm), which remains a priority for experiments in chemistry and life sciences. Apart from the superior spatial resolution of soft x-ray ptychography compared to STXM, samples with low absorption will also benefit from the high-quality phase imaging that the technique offers.

Concerning the achievable spatial resolution with soft x-ray ptychography, approaching the detector closer to the sample or larger area detectors in the future will improve the resolution toward the diffraction limit.

## ACKNOWLEDGMENTS

Development of SOPHIE was supported by the Swiss Nanoscience Institute (SNI). Soft x-ray ptychography measurements were performed at the Surface/Interface Microscopy (SIM-X11MA) beamline of the Swiss Light Source (SLS), Paul Scherrer Institut, Villigen, Switzerland and the SoftiMAX beamline of the MAX IV Laboratory. Research conducted at MAX IV, a Swedish national user facility, was supported by the Swedish Research Council under Contract No. 2018-07152, the Swedish Governmental Agency for Innovation Systems under Contract No. 2018-04969, and Formas under Contract No. 2019-02496. T.A.B. acknowledges funding from SNI and the European Regional Development Fund (ERDF). N.W.P. received funding from the European Union's Horizon 2020 research and innovation programme under Marie Skłodowska-Curie Grant Agreement No. 884104. We thank I. Beinik and K. Thånell for their support at the SoftiMAX beamline. We thank B. Rösner for the fabrication of the Siemens star. We thank E. Fröjdh, F. Baruffaldi, M. Carulla, J. Zhang, and A. Bergamaschi for the development of the iLGAD EIGER detector. The iLGAD sensors were fabricated at Fondazione Bruno Kessler (Trento, Italy).

## AUTHOR DECLARATIONS

### Conflict of Interest

The authors have no conflicts to disclose.

### Author Contributions

**Tim A. Butcher**: Data curation (equal); Formal analysis (equal); Investigation (equal); Methodology (equal); Software (equal); Validation (equal); Visualization (equal); Writing – original draft (equal); Writing – review & editing (equal). **Simone Finizio**: Data curation (equal); Formal analysis (equal); Investigation (equal); Methodology (equal); Resources (equal); Software (equal); Validation (equal); Writing – review & editing (equal). **Lars Heller**: Resources (equal); Visualization (equal). **Nicholas W. Phillips**: Data curation (equal); Formal analysis (equal); Investigation (equal); Software (equal); Writing – review & editing (equal). **Blagoj Sarafimov**: Resources (equal). **Carlos A. F. Vaz**: Resources (equal); Writing – review & editing (equal). **Armin Kleibert**: Funding acquisition (equal); Resources (equal); Writing – review & editing (equal). **Benjamin Watts**: Data curation (equal); Resources (equal); Software (equal); Visualization (equal); Writing – review & editing (equal). **Mirko Holler**: Resources (equal); Writing – review & editing (equal). **Jörg Raabe**: Conceptualization (equal); Data curation (equal); Funding acquisition (equal); Methodology (equal); Resources (equal); Software (equal); Supervision (equal); Writing – review & editing (equal).

## DATA AVAILABILITY

The data that support the findings of this study are available from the corresponding authors upon reasonable request.